\documentclass[prl,nofootinbib,twocolumn]{revtex4}

\usepackage{color}
\usepackage{graphicx}
\usepackage{amsmath}

\begin{document}

\title{Is there concordance within the concordance $\Lambda$CDM model?}

\author{Marco Raveri$^{1,2,3}$}

\smallskip
\affiliation{$^{1}$ SISSA - International School for Advanced Studies, Via Bonomea 265, 34136, Trieste, Italy \\
\smallskip
$^{2}$ INFN, Sezione di Trieste, Via Valerio 2, I-34127 Trieste, Italy \\
\smallskip
$^{3}$ INAF-Osservatorio Astronomico di Trieste, Via G.B. Tiepolo 11, I-34131 Trieste, Italy}

\begin{abstract}
We use a complete and rigorous statistical indicator to measure the level of concordance between cosmological data sets, without relying on the inspection of the marginal posterior distribution of some selected parameters.
We apply this test to state of the art cosmological data sets, to assess their agreement within the $\Lambda$CDM model.
We find that there is a good level of concordance between all the experiments with one noticeable exception.
There is substantial evidence of tension between the CMB temperature and polarization measurements of the {\it Planck} satellite and the data from the CFHTLenS weak lensing survey even when applying ultra conservative cuts.
These results robustly point toward the possibility of having unaccounted systematic effects in the data, an incomplete modelling of the cosmological predictions or hints toward new physical phenomena.
\end{abstract}

\maketitle

Our present understanding of the universe is based on the combination of several different cosmological observations that are joined in order to exploit their complementary sensitivities to distinct characteristics of our universe.
Several supernovae (SN) surveys are added together, in a single catalogue, to measure the expansion history at late times.
Different Baryon Acoustic Oscillations (BAO) surveys provide independent measurements of a cosmological standard ruler at several times. 
Large scale structure and weak lensing surveys measure the correlation of galaxies and weak lensing shear in many different redshift bins. These are combined together to get tomographic information on the clustering of cosmic structures. 
At last, measurements of the Cosmic Microwave Background (CMB) are reaching an extraordinary level of sensitivity, that allows to measure the CMB temperature fluctuations along with CMB polarization and lensing. 
These data are then joined together to exploit CMB sensitivity to both early and late times cosmology. \\
In the future, cosmological studies are going further in this direction. 
Wide large scale structure surveys, like Euclid~\cite{Euclid}, will combine maps of galaxies at several different redshifts, that will be joined with measurements of the CMB from the {\it Planck} satellite~\cite{Adam:2015rua} and sub-orbital experiments. \\
The observational efforts, that are driving cosmology toward a phase of extremely accurate, large scale, measurements, will all be joined together to learn all possible information about the initial conditions and the evolution of our universe.
In this program, however, a problem arises. \\
How can we be sure that the data sets, that we will be collecting, form a coherent picture, when interpreted within a model?
How do we quantify the agreement between them, to be aware of the possible presence of unaccounted systematic effects or hints toward new physical phenomena? \\
Testing the agreement between data sets, in a rigorous way, that goes beyond the comparison of the marginal distribution of some parameters, is critical in answering these questions.
The posterior of the model parameters is, in fact, not guaranteed to show tensions due to the marginalization procedure, that can alter discrepancies, that will be then misjudged.
Assessing whether the posterior distribution of two different data sets occupy a substantially different volume in the parameter space of a model is instead crucial as it could provide a useful guidance for the future research.
Answering these questions is also a useful sanity check for parameters estimation. 
The statistical inference on the parameters of a model should get stronger as we combine together different measurements and should not reflect the fact that we are joining low probability tails of the model posterior. \\
An estimate of the tensions between different data sets, based on the marginal posterior of cosmological parameters, has shown that indeed some discrepancies arise when combining several probes~\cite{Heymans:2013fya,Aubourg:2014yra,Ade:2015xua,Abbott:2015swa} that could point toward some extensions of the fiducial model~\cite{Wyman:2013lza,Hamann:2013iba,Costanzi:2014tna,Ade:2015rim, Hu:2015rva}.\\
In this {\it Letter} we briefly review how Bayesian inference can be used to answer quantitatively these questions and we comment on the advantages and possible drawbacks of this approach. 
We apply, for the first time, this statistical test to state of the art cosmological measurements and the $\Lambda$CDM model.
We report and interpret the results commenting on their relevance for future studies.

{\it The Data Concordance Test (DCT):} Bayesian statistics provides a clean way of dealing with the problem of combining data sets by means of hypotheses testing and in particular with its application to the problem of classification and decision making.
We have two data sets and we want to test whether we can describe them with the same set of parameters or not, within a given model. Based on the outcome of such operation we shall take a decision about combining them~\cite{Marshall:2004zd,Feroz:2008wr,March:2011rv,Amendola:2012wc,Verde:2013wza,Martin:2014lra,Karpenka:2014moa}. \\
Let us now consider two data sets $D_1$ and $D_2$ and a model $\mathcal{M}$. 
The two competing hypotheses that we want to compare are:
\begin{itemize}
\item $\mathcal{I}_0$: the two data sets can be characterized, within model $\mathcal{M}$, by the same (unknown) parameters; 
\item $\mathcal{I}_1$: the two data sets can be described, within model $\mathcal{M}$, with different (unknown) parameters.
\end{itemize}
Then, we compare the evidences for these two statements to obtain their odds-ratio. In particular by assigning non-committal priors for the two hypotheses we immediately have:
\begin{align} \label{Eq:DCTdef}
\mathcal{C}(D_1,D_2,\mathcal{M}) &= \frac{P(D_1\cup D_2|\,\mathcal{I}_0,\mathcal{M})}{P(D_1\cup D_2|\,\mathcal{I}_1,\mathcal{M})}  \nonumber \\
&= \frac{P(D_1\cup D_2|\mathcal{M})}{P(D_1|\mathcal{M})P(D_2|\mathcal{M})}
\end{align}
where $P(D_1\cup D_2|\mathcal{M})$ is the evidence of the joint data sets and $P(D_1|\mathcal{M})$ and $P(D_2|\mathcal{M})$ are the evidences of the single data sets. The last equality follows from the fact that under the hypothesis $\mathcal{I}_1$, the two data sets, $D_1$ and $D_2$, pertain to distinctly different classes and knowledge of one of them tells us nothing about the other. \\
We can interpret the odds resulting from this calculation with a classification scheme, like the Jeffreys' scale or others, depending on the decision that we have to perform afterwards.
In particular, when we have to decide if it is appropriate to combine two data sets, we can establish a threshold for the positive answer, based on the risk that we are willing to take, and act accordingly. A common choice~\cite{Marshall:2004zd} with this respect is to decide to follow the $\mathcal{I}_0$ hypothesis, combining the data, if $\log \mathcal{C} >0$ and $\mathcal{I}_1$ otherwise. \\
The DCT is relatively easy to compute, once we have at our disposal the tools to perform efficient evidence computations, and has some other advantages.
First of all the DCT is a quantitative and statistically rigorous prescription. 
It measures the odds, within model $\mathcal{M}$, of obtaining one data set given the other one.
Tensions between them are quantified in terms of odds of agreement or disagreement and are not based on the marginal distribution of the parameters. While the latter approach might point in the right direction if the likelihood is Gaussian, both in the data and the parameters, it might fail as soon as the posterior is slightly non-Gaussian.
Indeed, with the above assumptions, it can be shown~\cite{Marshall:2004zd,Feroz:2008wr} that the DCT reduces to the usual prescription for the marginal posterior of uncorrelated parameters.
When these requirements break down, however, the inspection of the parameters posterior becomes unreliable, in assessing tensions, as it tends to be biased.
In addition, as common sense suggests, the DCT naturally favours the combination of data sets, as long as there is no strong evidence that that should not be done~\cite{Marshall:2004zd,Amendola:2012wc}. The way in which this is automatically encoded in the computation of $\mathcal{C}$ is by weighting the prior volume with the likelihood volume, in a manner that resembles the Occam razor common to Bayesian model selection. If there is no clear indication on how to set the prior ranges, i.e. the previous knowledge of the model is vague, and the prior are consequently wide, the DCT favours the combination of data sets, as this might help in gaining knowledge of the model. Conversely if the priors are stronger than the data the DCT will disfavour the combination, as we already included in the prior choice the information that is coming from the combination of the data sets. \\
The DCT has also some disadvantages.
It does not give any indication whether the model is good by itself in fitting the data.
Being a comparative test, we can use it to judge if the agreement, within a given model, improves or not when combining two data sets but it is possible to have a model that fits very badly the data while the DCT might still favour their combination.
Another problem, that is particularly relevant when the DCT is used more than once on some data sets, is that it is not robust against over fitting. As immediately follows from the previous points, enlarging the parameter space with the introduction of an additional parameter will not decrease $\mathcal{C}$.
As a consequence, it is always possible to relax a tension between different measurements by introducing a new parameter, being it just a nuisance parameter, describing some systematic effects, or a parameter related to a different underlying physical modelling.
For this reason it is critical to use other statistical tools to assess whether the introduction of the additional parameter is really justified.
It is worth noticing, that as a by-product of the computation of $\mathcal{C}$, for the two different models, one has the relevant information to perform evidence based model comparison.
The last source of biases in the DCT are due to unaccounted correlations between the data sets. If two data sets are assumed to have independent errors and this is not the case, $\mathcal{C}$ will be biased toward positive values if the covariance between the errors of the two experiments is positive and toward negative values in the opposite case~\cite{Amendola:2012wc}.
\begin{figure*}[!t]
\centering
\includegraphics[width=1\textwidth]{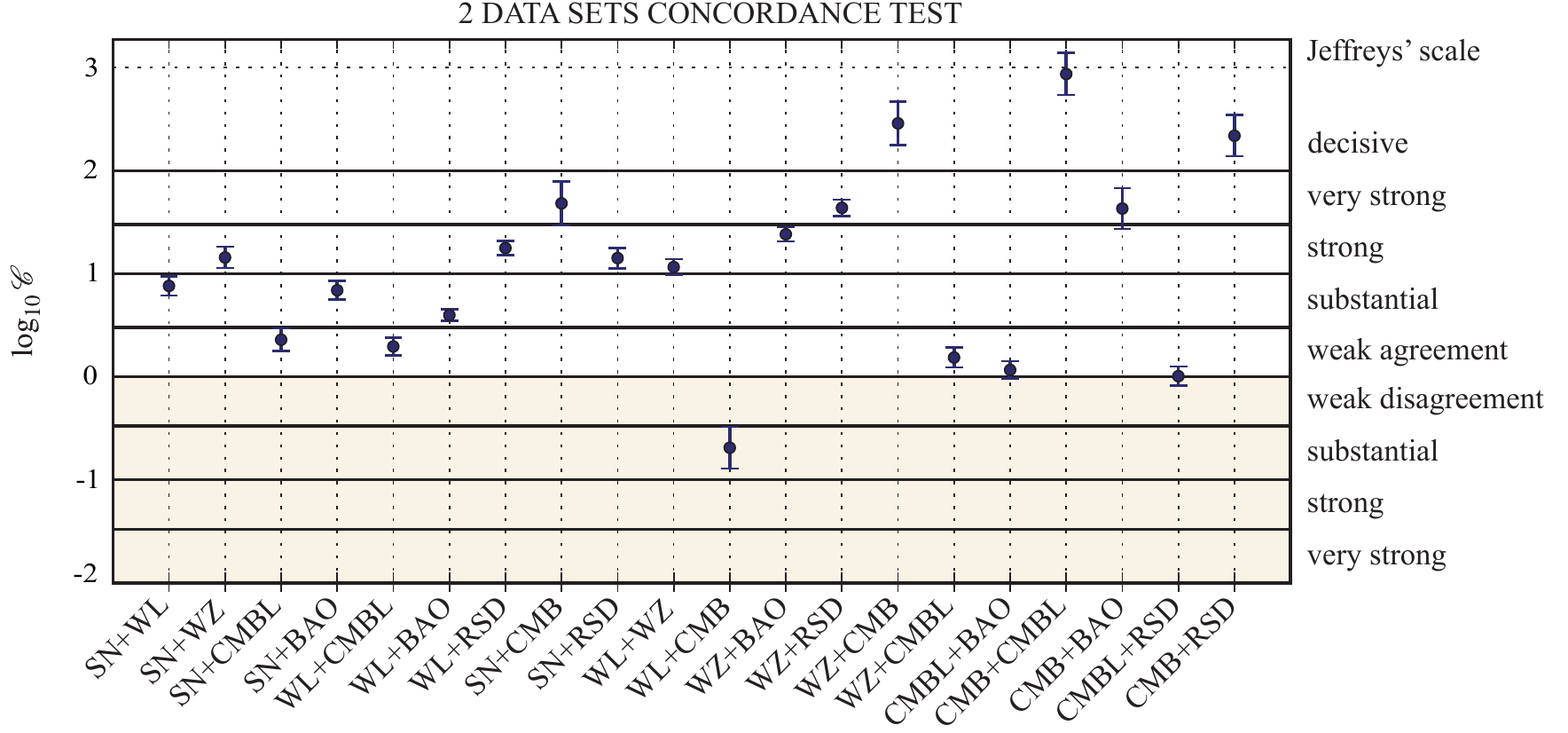}
\caption{The Data Concordance Test (DCT) performed on all the independent couples of the cosmological data sets described in the text. The shaded region highlights the values of $\mathcal{C}$ that point toward disagreement between data sets. The error bars represent the uncertainty associated with the nested sampling computation of the evidence.}
\label{Fig:DCT2D}
\end{figure*}

{\it Data sets and model:} we use several available cosmological data sets to perform a DCT over all the possible independent data couples, within the $\Lambda$CDM model. \\
The first data set that we consider consists of the ``Joint Light-curve Analysis'' (JLA) Supernovae sample, as introduced in~\cite{Betoule:2014frx}, which is constructed by the combination of the SNLS, SDSS and HST SNe data, together with several low redshift SNe. \\
We use the  WiggleZ Dark Energy Survey~\cite{wigz} measurements of the galaxy power spectrum as inferred from $170,352$ blue emission line galaxies over a volume of $1\,\mbox{Gpc}^3$~\cite{Drinkwater:2009sd,Parkinson:2012vd}. We marginalize over a scale independent linear galaxy bias for each of the four redshift bins, as in~\cite{Parkinson:2012vd}. \\
The third data set that we examine consists of the measurements of the galaxy weak lensing shear correlation function as provided by the 
Canada-France-Hawaii Telescope Lensing Survey (CFHTLenS)~\cite{Heymans:2013fya}. This is a $154$ square degree multi-colour survey, optimised for weak lensing analyses, that spans redshifts ranging from $z\sim0.2$ to $z\sim 1.3$.
Here we consider the data subdivided into $6$ redshift bins and we applied ultra-conservative cuts, as in~\cite{Ade:2015rim}, that exclude $\xi_{-}$ completely and cut the $\xi_+$ measurements at scales smaller than $\theta=17'$ for all the tomographic redshift bins.
As discussed in~\cite{Ade:2015rim}, these cuts make the CFHTLenS data insensitive to the modelling of the non-linear evolution of the power spectrum. \\
We include in this study the measurements of the CMB fluctuations in both temperature and polarization as released by the {\it Planck} satellite~\cite{Ade:2015xua, Aghanim:2015xee}. 
At large angular scales the {\it Planck} release implements a joint pixel-based likelihood including both temperature and E-B mode polarization 
for the multipoles range of $\ell \leq 29$, as described in~\cite{Aghanim:2015xee}.
At smaller angular scales we use the \texttt{Plik} likelihood~\cite{Aghanim:2015xee} for CMB measurements of the TT, TE and EE power spectra, as extracted from the $100$, $143$, and $217$ GHz HFI channels. 
We refer to the combination of the low-$\ell$ TEB measurements and the high-$\ell$ TT TE EE data as the CMB compilation. \\
We also include in the analysis the {\it Planck} 2015 full-sky lensing potential power spectrum~\cite{Ade:2015zua} in the multipoles range $4\leq \ell \leq 400$ as obtained with the SMICA code, hereafter called CMBL. \\
We also employ BAO measurements of: the SDSS Main Galaxy Sample at $z_{\rm eff}=0.15$~\cite{Ross:2014qpa}; the BOSS DR11 ``LOWZ" sample at $z_{\rm eff}=0.32$~\cite{Anderson:2013zyy}; the BOSS DR11 CMASS at $z_{\rm eff}=0.57$ of~\cite{Anderson:2013zyy}; and the 6dFGS survey at $z_{\rm eff}=0.106$~\cite{Beutler:2011hx}, all joined together in the data set that we dub BAO. \\
In addition we consider the redshift space distortion (RSD) measurements of BOSS CMASS-DR11 as analysed in~\cite{Beutler:2013yhm} and~\cite{Samushia:2013yga}. When these data are used we exclude the BOSS-CMASS results of~\cite{Anderson:2013zyy} from the BAO likelihood to avoid double counting.
We refer to the data set obtained by combining RSD measurements and the independent BAO measurements as the RSD one.\\
By means of the DCT we perform a test of the data concordance within the $6$ parameter $\Lambda$CDM model.
To compute non-linear corrections to the matter power spectrum and the lensed CMB power spectra, we use the halofit approach~\cite{Smith:2002dz} with the updates of~\cite{Takahashi:2012em}.
We use the CAMB code~\cite{CAMB,Lewis:1999bs} to compute the predictions for all cosmological observables of interest and we use the likelihoods of the previously described data sets, as implemented in CosmoMC~\cite{Lewis:2002ah}.
We compute the evidence by means of the nested sampling algorithm and its implementation in the PolyChord code~\cite{Handley:2015fda,polychord}.
In order to assess the agreement between the above data sets, in the set-up commonly used for parameter estimation, we use the standard CosmoMC prior on the $\Lambda$CDM model parameters. \\

\begin{figure}[!]
\centering
\includegraphics{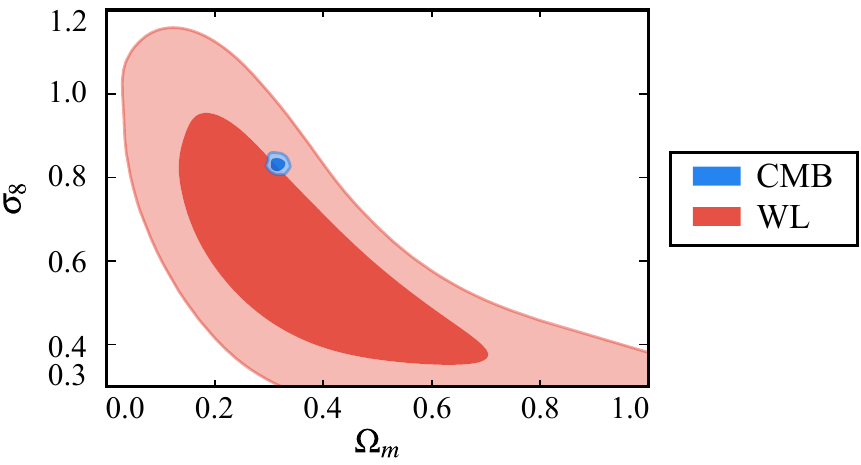}
\caption{The marginalized joint posterior for $\Omega_m$ and the amplitude of the linear power spectrum on the scale of $8\,h^{-1}\,\mbox{Mpc}$ for different data sets, as shown in legend. The darker and lighter shades correspond respectively to the $68\%$ C.L. and the $95\%$ C.L. regions.}
\label{Fig:marginal_large}
\end{figure}
{\it Results:} the results of the DCT of all the independent couples of the data sets described above is shown in figure~\ref{Fig:DCT2D}.
We can see that the combination of CMB and WL data shows evidence of substantial disagreement. 
It is worth noticing that the marginal distribution of the parameters is not displaying strong discrepancies. 
In figure~\ref{Fig:marginal_large} we show the joint marginalized posterior of the parameters $\sigma_8$ and $\Omega_m$ that is commonly used~\cite{Ade:2015xua,MacCrann:2014wfa,Liu:2015xfa,Abbott:2015swa} to discuss tensions between these kind of data sets.
As we can see, the constraints coming from the two data sets seem consistent at $68\%$ C.L. as in~\cite{Ade:2015xua,Ade:2015rim}. 
This is a clear example where marginalising over a high dimensional non-Gaussian likelihood to get the posterior of some parameters biases the conclusions on the possible tensions between data sets.
It is clear, from this study, that the DCT helps in assessing whether discrepancies, over the whole parameter space of a model, are statistically relevant and require further investigation. \\
The $\mathcal{C}$ values involving CMB lensing (CMBL) data are all weakly pointing toward agreement.
CMBL+BAO and CMBL+RSD, in particular, are borderline between agreement and disagreement. For both data sets, this comes from some discrepancies in the determination of the background parameters. RSD data, in addition, are also penalised by some discrepancy in the determination of the amplitude of scalar perturbations.
The results of the DCT involving CMB and SN, BAO, RSD and WZ are on the high end of the comparison scale, with values that range from very strong to decisive. This is largely expected as we are combining a probe that is extremely sensitive to all the cosmological parameters (CMB) with other data that probe only a sub-set of them. \\
Surprisingly enough, when combining CMB and CMBL the DCT reaches a very high value, the maximum achieved in this comparison. 
This seems suspicious, as CMBL was found in weak agreement with all other data sets while CMB was displaying a good agreement. 
It is beyond the scope of this work to investigate the possible causes of this behaviour. We can, however, have some ideas on its origin from the properties of the DCT discussed before. 
It seems unlikely that this discrepancy arises due to a factor involving the likelihood volume because its shrinking, with respect to the prior volume, is mainly CMB driven.
It seems also improbable that this is due to the difference in the best fit $\chi^2$ because that is again CMB dominated.
A plausible reason for this behaviour could be data set dependencies. We know that CMB and lensing measurements are correlated and, in particular, there is a positive cross-correlation between CMB temperature fluctuations and the CMB lensing spectrum. 
Even if the two seem to agree very well, neglecting this correlation can bias toward positive values $\mathcal{C}$ and, as shown here, it is surely an effect that requires a deeper investigation. \\
SN data show good agreement with BAO and RSD measurements, from substantial to strong on a Jeffreys' scale, as they agree on the determination of the parameters describing background evolution. The agreement between SN and RSD is slightly higher than BAO as this data set is also sensitive to some perturbations parameters.
Agreement between WL and SN, BAO and RSD is also good as the DCT is rewarding the additional leverage on perturbations parameters that comes from WL measurements. 
For the same reason, a good agreement is found also for WZ and SN, BAO and RSD. Noticeably the values of $\mathcal{C}$ are slightly higher than the previous ones. This reflects the fact that, due to the presence of non-linear scales in WZ data, the constraints on perturbations parameters are stronger than the previous ones.
Testing the combination of WL and WZ data then results in strong agreement. The two data roughly agree on the background parameters and the additional constraining power of WZ on perturbation parameters favours the combination of these two data sets.

In conclusion, we have used Bayesian hypothesis testing to assess quantitatively whether there is concordance, within the $\Lambda$CDM model, between several different cosmological experiments.
This test, that we dubbed DCT, allows to compute the odds that two data sets can be described by the same choice of cosmological parameters and thus gives a way of measuring the statistical significance of tensions between different measurements.
We have commented some of the properties that make this test a reliable tool that extends, with statistical rigour, other commonly used approaches.
We applied this test to the combinations of some of the most relevant cosmological data sets to date and found, overall, a good agreement between geometrical probes and other perturbations measurements.
We showed, however, that the lensing of the CMB is only weakly in agreement with all other cosmological data sets but CMB itself.
The odds of this agreement are suspiciously high, given the other results, and might point toward the necessity of including the cross correlation between CMB lensing and CMB spectra into future studies.
At last, we found substantial evidence for a disagreement between WL data of CFHTLenS and CMB measurements of {\it Planck}.
We showed that a similar conclusion would not be drawn by inspecting the marginal posterior of some parameters. \\
The investigation of the tensions found in this work, and how they are relieved in extended models, is a primary goal as these could point toward the presence of unaccounted systematic effects, an incomplete modelling of the cosmological predictions or the presence of new physical phenomena.

\vskip 10pt
{\it Acknowledgements:}
I am grateful to Carlo Baccigalupi, Nicola Bartolo, Stefano Camera, Noemi Frusciante, Alan Heavens, Bin Hu, Michele Liguori, Matteo Martinelli, Sabino Matarrese, Levon Pogosian, Alessandra Silvestri, Licia Verde and Matteo Viel for useful discussions and comments on the manuscript.
I acknowledge partial support from the INFN-INDARK initiative. 
I thank the Instituut Lorentz (Leiden University) for the computing resources made available for conducting the research reported in this {\it Letter}.

\end{document}